\documentclass[12pt]{article}
\usepackage{amsmath} 
\input{epsf} 
\def\ltap{\raisebox{-.6ex}{\rlap{$\,\sim\,$}} \raisebox{.4ex}{$\,<\,$}}

\newcommand\as{\alpha_{\mathrm{S}}} 
\newcommand\f[2]{\frac{#1}{#2}} 
\def\beq{\begin{equation}} 
\def\eeq{\end{equation}} 
\def\beeq{\begin{eqnarray}} 
\def\eeeq{\end{eqnarray}} 
 
\def\to{\rightarrow}

\def\ptmin{p_{T{\rm min}}}
\def\ptmax{p_{T{\rm max}}}

\begin{document} 

\begin{titlepage}
\renewcommand{\thefootnote}{\fnsymbol{footnote}}
\begin{flushright}
hep-ph/0703012
     \end{flushright}
\par \vspace{10mm}

\begin{center}
{\Large \bf An NNLO subtraction formalism in hadron collisions and}
\\[0.5cm]
{\Large \bf its application to Higgs boson production at the LHC}
\end{center}
\par \vspace{2mm}
\begin{center}
{\bf Stefano Catani}~~and~~{\bf Massimiliano Grazzini}\\

\vspace{5mm}

INFN, Sezione di Firenze\\ and Dipartimento di Fisica,
Universit\`a di Firenze,\\ 
I-50019 Sesto Fiorentino, Florence, Italy\\

\vspace{5mm}

\end{center}

\par \vspace{2mm}
\begin{center} {\large \bf Abstract} \end{center}
\begin{quote}
\pretolerance 10000

We consider higher-order QCD corrections to the production
of colourless high-mass systems
(lepton pairs, vector bosons, Higgs bosons, 
$\dots$)
in
hadron collisions.
We propose a new formulation of the subtraction method 
to numerically compute arbitrary
infrared-safe observables for this class of processes. 
To cancel the infrared divergences, we
exploit the universal
behaviour of the associated transverse-momentum ($q_T$) distributions in the
small-$q_T$ region. The method is illustrated in general terms up to the
next-to-next-to-leading order (NNLO) in QCD perturbation theory.
As a first explicit application, we study Higgs boson production
through gluon fusion. Our calculation is implemented in 
a parton level Monte Carlo program that includes
the decay of the Higgs boson in two photons. We present 
selected numerical results
at the LHC.

\end{quote}

\vspace*{\fill}
\begin{flushleft}
March 2007

\end{flushleft}
\end{titlepage}

\setcounter{footnote}{1}
\renewcommand{\thefootnote}{\fnsymbol{footnote}}

The dynamics of scattering processes at high-momentum scales, $Q$,
is
well described by perturbative QCD.
Thanks to asymptotic freedom, the cross section for sufficiently 
inclusive reactions can be computed as a series expansion
in the QCD coupling $\as(Q^2)$.
Until few years ago, the standard for
such calculations was next-to-leading order (NLO) accuracy.
Next-to-next-to-leading order (NNLO)
results were known only for 
few
highly-inclusive reactions
(see e.g. Refs.~\cite{sigmatot,Hamberg:1990np,Higgstot}).

The extension from NLO to NNLO accuracy is desirable
to improve the QCD predictions and to better assess
their uncertainties. In particular, this extension is certainly important in 
two cases: in those processes whose NLO corrections are comparable to the
leading order (LO) contribution; in those `benchmark' processes that are
measured with high experimental precision.
Such a task, however, implies finding methods and techniques to practically
achieve the cancellation of infrared (IR) divergences that appear at 
intermediate steps of the calculations.

Recently, a new general method \cite{Anastasiou:2003gr},
based on sector decomposition \cite{sector}, has been proposed
and applied to the NNLO QCD calculations of $e^+e^-\to 2~{\rm jets}$
\cite{Anastasiou:2004qd}, Higgs \cite{Hdiff}
and vector \cite{DYdiff} boson production in hadron collisions, and to the NNLO
QED calculation of the electron energy spectrum in muon decay
\cite{Anastasiou:2005pn}.
The calculations of Refs.~\cite{Hdiff,DYdiff} are encoded in publicly available
numerical programs that
allow the user to compute the corresponding cross sections with arbitrary 
cuts on the momenta of the partons produced in the final state.

The traditional approach to
perform NLO computations
is based on the introduction of auxiliary cross sections that are obtained
by approximating the QCD scattering amplitudes in the relevant IR (soft and
collinear) limits.
This strategy led to the proposal
of the {\em subtraction} \cite{Ellis:1980wv}
and {\em slicing} \cite{Fabricius:1981sx} methods.
Exploiting the universality properties of soft and collinear emission,
these methods were later developed in the form of 
general algorithms \cite{Giele:1991vf,Frixione:1995ms,Catani:1996vz}.
These algorithms make possible
to perform NLO calculations in a (relatively) straightforward manner,
as soon as the corresponding QCD amplitudes are available.
In recent years, several research groups have been working on general NNLO
extensions of the subtraction method 
\cite{Kosower,Weinzierl,Frixione:2004is,GGG,ST}.
Although
NNLO results have been obtained only in some specific processes
($e^+e^-\to 2~{\rm jets}$ \cite{Gehrmann-DeRidder:2004tv,Weinzierl:2006ij}
and, partly, $e^+e^-\to 3~{\rm jets}$ \cite{Gehrmann-DeRidder:2004xe}),
in the case of lepton collisions
some of these general projects are near to completion. 

In this letter we reconsider the problem of the extension of the
subtraction method to NNLO. Rather than aiming at a general formulation,
we limit ourselves to considering a specific, though important, class of 
processes: the production of colourless high-mass systems in hadron collisions.
We present a formulation of the subtraction method for this class
of processes, and we apply
it to the NNLO calculation of Higgs boson production via the gluon fusion
subprocess $gg\to H$. This explicit application crosschecks 
the results of Ref.~\cite{Hdiff}, by using a completely independent method.

We consider the inclusive hard-scattering reaction
\begin{equation}
h_1+h_2\to F(Q)+X,
\end{equation}
where the collision of the two hadrons $h_1$ and $h_2$
produces the triggered final state $F$. The final 
state $F$ consists of one or more colourless particles (leptons, photons, vector
bosons, Higgs bosons, $\dots$) with momenta $q_i$ and total invariant mass $Q$
($Q^2=(\sum_i q_i)^2$).
Note that, since $F$ is colourless, the LO partonic subprocess
is either $q{\bar q}$ annihilation, as in the case of the Drell--Yan process, 
or $gg$ fusion,
as in the case of Higgs boson production.

At NLO, two kinds of corrections contribute: i) {\em real} corrections,
where one parton recoils against $F$; ii) {\em one-loop virtual} corrections to
the LO subprocess. Both contributions are separately 
IR divergent, but 
the divergences cancel in the sum.
At NNLO, three kinds of corrections must be considered: i) {\em double real} 
contributions, where two partons recoil against $F$; ii) {\em real-virtual} 
corrections, where one parton recoils against $F$ at one-loop order; 
iii) {\em two-loop virtual} corrections to the LO subprocess.
The three contributions are still
separately divergent, and the calculation has
to be organized so as to explicitly achieve the cancellation of the 
IR divergences.

Our
method
is based on a (process- and observable-independent) generalization
of the
procedure
used in the specific NNLO calculation of
Ref.~\cite{Catani:2001cr}.
We first note that, at LO, the transverse momentum 
${\bf q}_{\, T}= \sum_i {\bf q}_{\, Ti}$ of the triggered final state $F$ is exactly zero.
As a consequence, as long as $q_T\neq 0$, the (N)NLO contributions are actually given by the (N)LO 
contributions to the triggered final state $F+{\rm jet(s)}$.
Thus, we can write the 
cross section as
\begin{equation}
\label{Fplusjets}
d{\sigma}^{F}_{(N)NLO}|_{q_T\neq 0}=d{\sigma}^{F+{\rm jets}}_{(N)LO}
\;\; .
\end{equation}
This means that, when $q_T\neq 0$, the 
IR divergences in our NNLO calculation are those in 
$d{\sigma}^{F+{\rm jets}}_{NLO}$: they can be handled and
cancelled by using 
available NLO formulations of the subtraction method.
The only remaining singularities of NNLO type are associated to the limit 
$q_T\to 0$, and we treat them by an additional subtraction. 
Our key point is that the singular behaviour 
of $d{\sigma}^{F+{\rm jets}}_{(N)LO}$ when $q_T\to 0$ is well known:
it comes out in the
resummation program \cite{Catani:2000jh}
of logarithmically-enhanced contributions
to transverse-momentum distributions.
Then,
to perform the additional subtraction, we follow the formalism used in 
Ref.~\cite{qtresum,ww} to combine resummed and fixed-order calculations.

The following sketchy presentation is illustrative; 
the details will appear elsewhere.
We use a shorthand notation that mimics the notation 
of Ref.~\cite{qtresum}. We define the subtraction 
counterterm\footnote{The symbol $\otimes$ understands convolutions over momentum
fractions and sum over flavour indeces of the partons.}
\begin{equation}
\label{ct}
d{\sigma}^{CT}
=
d{\sigma}_{LO}^F\otimes\Sigma^F(q_T/Q)\, d^2{\bf q}_{\, T}.
\end{equation}
The function $\Sigma^F(q_T/Q)$ embodies the
singular behaviour of $d{\sigma}^{F+{\rm jets}}$
when $q_T\to 0$. In this limit it can be expressed as
follows in terms of $q_T$-independent coefficients $\Sigma^{F(n;k)}$:
\begin{equation}
\label{sigmalimit}
\Sigma^F(q_T/Q)
\xrightarrow[q_T\to 0]{}
\sum_{n=1}^\infty
\left(\f{\as}{\pi}\right)^n\sum_{k=1}^{2n}
\Sigma^{F(n;k)} \;\f{Q^2}{q_T^2}\ln^{k-1} \f{Q^2}{q_T^2}  \;\; .
\end{equation}
The extension of Eq.~(\ref{Fplusjets}) to include
the contribution at $q_T=0$ is finally:
\begin{equation}
\label{main}
d{\sigma}^{F}_{(N)NLO}={\cal H}^{F}_{(N)NLO}\otimes d{\sigma}^{F}_{LO}
+\left[ d{\sigma}^{F+{\rm jets}}_{(N)LO}-
d{\sigma}^{CT}_{(N)LO}\right]\;\; .
\end{equation}
Comparing with the right-hand side of
Eq.~(\ref{Fplusjets}), we have subtracted
the truncation of Eq.~(\ref{ct}) at (N)LO
and added a contribution at $q_T=0$ needed to 
obtain the correct total cross section.
The coefficient ${\cal H}^{F}_{(N)NLO}$ does not depend on $q_T$
and is obtained by the (N)NLO truncation of the perturbative function
\begin{equation}
{\cal H}^{F}=1+\f{\as}{\pi}\,
{\cal H}^{F(1)}+\left(\f{\as}{\pi}\right)^2
{\cal H}^{F(2)}+ \dots \;\;.
\end{equation}
A few comments are in order.
\vspace*{-5mm}
\begin{itemize}
\item The counterterm of Eq.~(\ref{ct})
regularizes the singularity of $d{\sigma}^{F+{\rm jets}}$ when $q_T\to 0$:
the term in the square bracket on the right-hand side of Eq.~(\ref{main}) is 
thus IR finite (or, better, integrable over $q_T$). Note that, at NNLO,
$d{\sigma}^{CT}_{(N)LO}$ acts as a counterterm for the {\em sum} of the two
contributions to $d{\sigma}^{F+{\rm jets}}$: the 
{\em double real} plus {\em real-virtual} contributions. Once 
$d{\sigma}^{F+{\rm jets}}$ has generated a weighted `event', 
$d{\sigma}^{CT}_{(N)LO}$ generates a corresponding counterevent with
LO kinematics (i.e. with $q_T=0$) and with weight $\Sigma^F(q_T/Q)$, where $q_T$
is the transverse momentum of $F$ in the `event'.
\item The explicit form of the counterterm in Eq.~(\ref{ct}) has some degrees of
arbitrariness. 
The LO kinematics of the counterevent can be defined by absorbing
in different ways the $q_T$-recoil of the `event': the only constraint is that
the `event' kinematics smoothly approaches the counterevent kinematics when
$q_T \to 0$. The counterterm function $\Sigma^F(q_T/Q)$ can be defined 
in different ways: the key property is that, in the small-$q_T$ limit, it 
must have the form given in Eq.~(\ref{sigmalimit}).
Note that the perturbative coefficients $\Sigma^{F(n;k)}$ are 
universal\footnote{More precisely, the NNLO coefficients $\Sigma^{F(2;1)}$
and $\Sigma^{F(2;2)}$
have a non-universal contribution that, nonetheless, is proportional
to the NLO coefficient ${\cal H}^{F(1)}$.}:
they only depend on the type of partons (quarks or gluon) involved in the
LO partonic subprocess ($q{\bar q}$ annihilation or $gg$ fusion).
\item The simplicity of the LO subprocess is such that final-state partons 
actually appear only in 
the term $d{\sigma}^{F+{\rm jets}}$ on the
right-hand side of Eq.~(\ref{main}).
Therefore,
arbitrary IR-safe cuts on the jets at (N)NLO can effectively be accounted for 
through a (N)LO computation. Owing to this feature,
our NNLO extension of the subtraction formalism is observable-independent.
\item At NLO (NNLO), the physical information of the {\em one-loop (two-loop)
virtual} correction to the LO subprocess is 
contained in the coefficients 
${\cal H}^{(1)}$ (${\cal H}^{(2)}$). 
Once an explicit form of Eq.~(\ref{ct}) is chosen, 
the hard coefficients ${\cal H}^{F (n)}$ are uniquely 
identified (a different choice would correspond to different ${\cal H}^{F (n)}$).
\end{itemize}

According to Eq.~(\ref{main}), the NLO calculation  of $d{\sigma}^{F}$ 
requires the knowledge
of ${\cal H}^{F(1)}$ and the LO calculation of $d{\sigma}^{F+{\rm jets}}$.
The general (process-independent) form of 
the coefficient 
${\cal H}^{F(1)}$ is basically known: the precise relation between 
${\cal H}^{F(1)}$ and the IR finite part of the
one-loop correction to a generic LO subprocess is explicitly derived in 
Ref.~\cite{deFlorian:2000pr}.

At NNLO, the coefficient ${\cal H}^{F(2)}$ is also needed, together with the
NLO calculation of $d{\sigma}^{F+{\rm jets}}$.
Although the general structure\footnote{It could be derived by extending the
${\cal O}(\as^2)$ calculation of Ref.~\cite{deFlorian:2000pr} to compute
subleading logarithms. Work along these lines is under way.}
of the coefficients ${\cal H}^{F(2)}$
is presently unknown,
we have 
completed the calculation of 
${\cal H}^{H(2)}$ for Higgs boson production in
the large-$M_{top}$ limit.
Since the NLO corrections to $gg\to H+{\rm jet(s)}$ are available 
\cite{deFlorian:1999zd} 
in the same limit,
we are able to present a first application of Eq.~(\ref{main})
at NNLO. We have 
encoded
our computation in a parton level
Monte Carlo program, in which
we can implement arbitrary IR-safe cuts on the final state.

In the following
we present numerical results for Higgs boson production at the LHC.
We use the MRST2004 parton distributions \cite{Martin:2004ir},
with densities and $\as$ evaluated at each corresponding order
(i.e., we use $(n+1)$-loop $\as$ at N$^n$LO, with $n=0,1,2$). The 
renormalization and factorization scales are fixed to the value 
$\mu_R=\mu_F=M_H$, where $M_H$ is the mass of the Higgs boson.

\begin{figure}[htb]
\begin{center}
\begin{tabular}{c}
\epsfxsize=10truecm
\epsffile{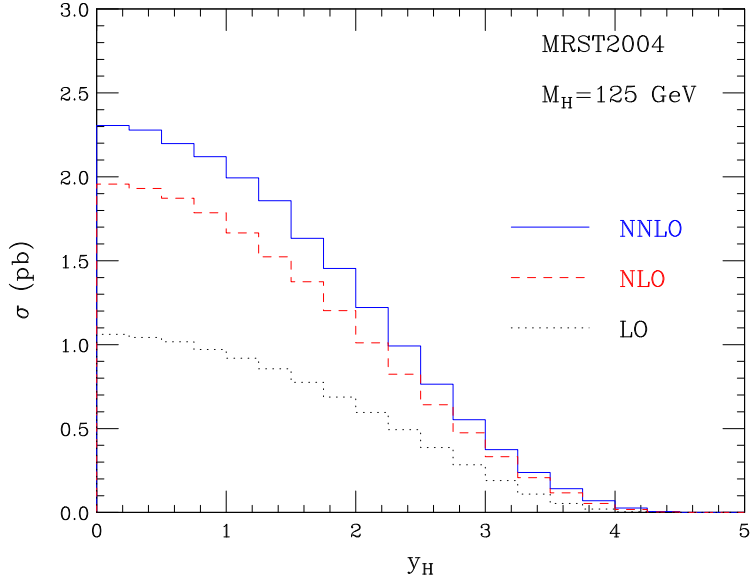}\\
\end{tabular}
\end{center}
\caption{\label{fig:incl}
{\em Bin-integrated rapidity distribution of the Higgs boson with 
$M_H=125$~GeV: results at LO (dotted), NLO (dashed) and NNLO (solid).
 }}
\end{figure}
In Fig.~\ref{fig:incl} we consider $M_H=125$~GeV, and we show the 
bin-integrated 
rapidity distribution of the
Higgs boson at LO (dotted), NLO (dashed) and NNLO (solid).
\begin{figure}[htb]
\begin{center}
\begin{tabular}{c}
\epsfxsize=10truecm
\epsffile{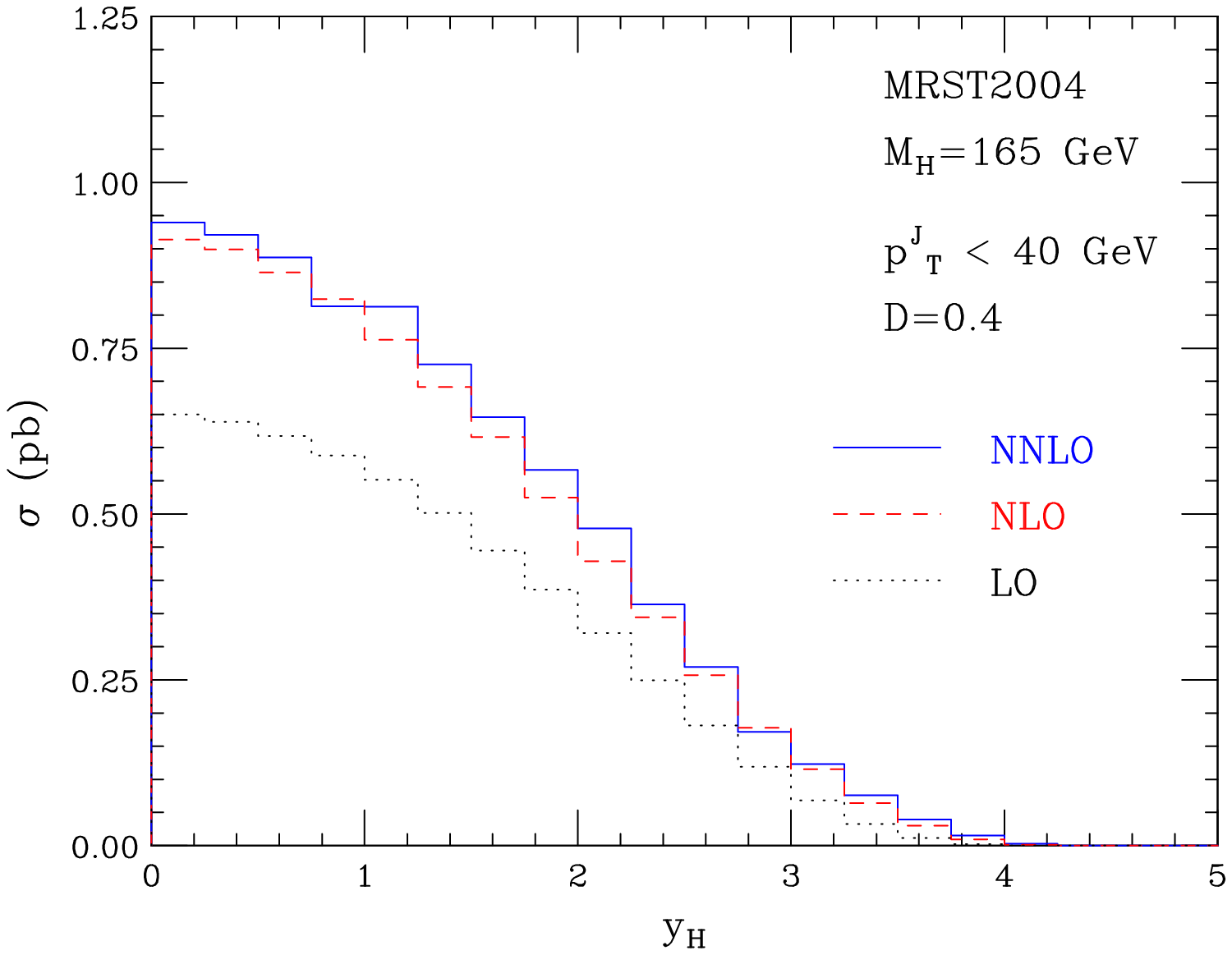}\\
\end{tabular}
\end{center}
\caption{\label{fig:veto}
{\em Bin-integrated rapidity distribution of the Higgs boson with 
$M_H=165$~GeV. Final-state jets
are required to have transverse momentum smaller than $40$~GeV.}}
\end{figure}
The impact of the NNLO corrections on the NLO result
is mildly dependent on the rapidity $y_H$ when $|y_H|\ltap 3$.
The total cross section increases by about $19\%$
when going from NLO to NNLO.

When searching for the Higgs boson 
in the $H\to WW$ channel, a jet veto is typically required 
to suppress the $WW$ background from  $t{\bar t}$ production.
In Fig.~\ref{fig:veto} we present the rapidity distribution of the Higgs boson
with $M_H=165$~GeV. In this case we apply a veto on the jets that recoil
against the Higgs boson.
Jets are reconstructed by using the $k_T$ algorithm \cite{ktalg} 
with jet size $D=0.4$
\footnote{In our calculation up to NLO,
the $k_T$ algorithm and the 
cone algorithm \cite{Blazey:2000qt} are equivalent. At NNLO, 
the $k_T$ algorithm is equivalent to the 
cone algorithm (with cone size $R=D$) {\em without} midpoint seeds, while the
cone algorithm {\em with} midpoint seeds would lead to (slightly) different
results.
The cone algorithm {\em without} midpoint seeds
would be infrared unsafe starting from N$^3$LO.};
each jet is
required to have transverse momentum smaller than $40$~GeV\footnote{At NNLO,
a jet may consist of two partons. In this case, the transverse momentum
of the jet is the vector sum of the transverse momenta of the two partons.}.
As is known \cite{Catani:2001cr,Hdiff}, the impact of higher-order corrections 
is reduced when a jet veto is applied. In the present case,
the impact of the NNLO corrections on the NLO total cross section
is reduced from 20 to $5~\%$.
\begin{figure}[htb]
\begin{center}
\begin{tabular}{c}
\epsfxsize=12truecm
\epsffile{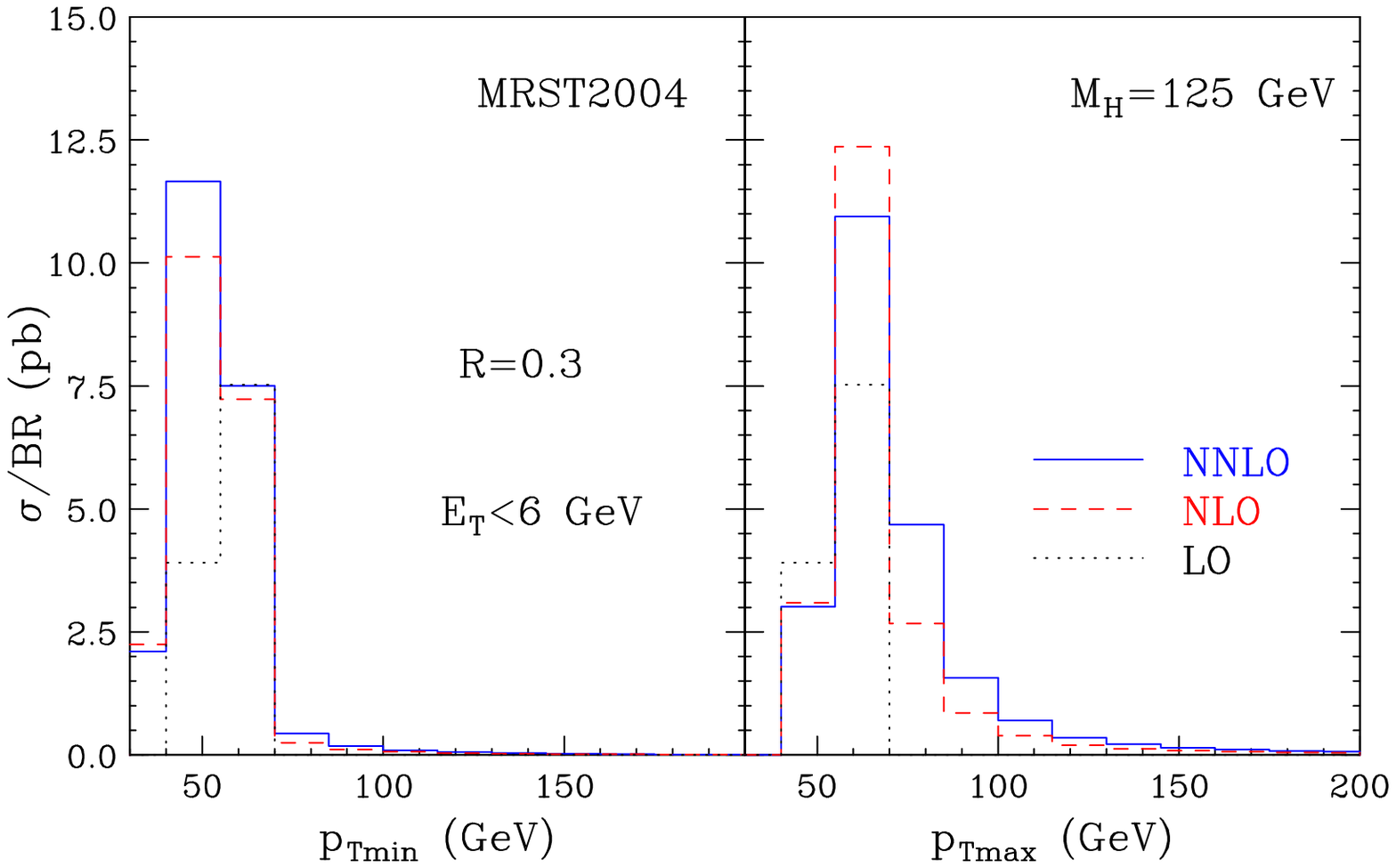}\\
\end{tabular}
\end{center}
\caption{\label{fig:isol}
{\em Distributions in $\ptmin$ and $\ptmax$ for the diphoton signal at 
the LHC. The cross section is divided by the branching ratio in two photons.}}
\end{figure}

We finally consider the Higgs boson decay in the
$H\to \gamma\gamma$ channel and follow
Ref.~\cite{CMStdr} to apply 
cuts on the photons.
For each event, we classify the photon transverse momenta according to their
minimum and maximum value,  
$\ptmin$ and $\ptmax$. The photons are required to be in the
central rapidity region, $|\eta|<2.5$, with  $\ptmin>35$~GeV
and $\ptmax>40$~GeV. We also require the photons to be isolated:
the hadronic (partonic) transverse energy in a cone of radius $R=0.3$ along the
photon direction 
has to be smaller 
than 6~GeV. When $M_H=125$~GeV, by applying these cuts
the impact of the NNLO corrections on the NLO total cross section
is reduced from 19\% to 11\%.
 
In Fig.~\ref{fig:isol} we plot
the 
distributions in $\ptmin$ and $\ptmax$
for the $gg\to H\to\gamma\gamma$ signal.
We note that the shape of these distributions sizeably
differs when going from LO to NLO and to NNLO.
The origin of these perturbative instabilities is well known 
\cite{Catani:1997xc}.
Since the LO spectra
are kinematically bounded by $p_T\leq M_H/2$,
each higher-order perturbative contribution produces
(integrable) logarithmic singularities in the vicinity of
that boundary. More detailed studies are necessary to assess
the theoretical uncertainties of these fixed-order results
and the relevance of all-order resummed calculations.
A similar comment applies to the distribution of the variable 
$(\ptmin+\ptmax)/2$, 
which is computed, for instance, in Refs.~\cite{Hdiff,Stockli:2005hz}.

We have illustrated 
an extension of the subtraction formalism to
compute NNLO QCD corrections to
the production of high-mass systems in hadron collisions. We have
considered an explicit application of
our method to the NNLO computation of $gg\to H\to\gamma\gamma$
at the LHC, and we have presented few selected results, including
kinematical
cuts
on the final state. The computation parallels the one of
Ref.~\cite{Hdiff}, but it is performed with a completely independent method. 
In the quantitative studies that we have carried out, the two computations
give results in numerical agreement. 
In our approach
the calculation is
directly implemented in a parton level event generator.
This feature makes it particularly suitable for practical applications
to the computation of distributions in the form of bin histograms.
We plan to release a public version of our program in the near future.
We also plan to apply the method to other hard-scattering processes.

\vspace*{0.5cm}

\noindent {\bf  Acknowledgements.} 
We would like to thank Daniel de Florian for helpful discussions and comments.

\end{document}